\documentclass[a4paper]{jpconf}

\usepackage{graphicx}
\usepackage{cite}
\usepackage{amsmath}
\usepackage{iopams}
\usepackage{bm}

\newcommand* {\vek}[1]{{\ensuremath{\bm{\mathrm{#1}}}}}
\newcommand* {\vekc}[1]{{\ensuremath{\bm{\mathcal{#1}}}}}
\newcommand* {\vekhc}[1]{{\ensuremath{\bm{\hat{\mathcal{#1}}}}}}
\newcommand* {\kk}{\vek{k}}
\newcommand* {\ee}{\ensuremath{\mathrm{e}}}
\newcommand* {\ubar}[1]{\text{\b{$#1$}}}

\DeclareMathOperator{\arccot}{arccot}

\begin{document}

\title{Longitudinal magnetoconductivity and magnetodielectric effect
  in bilayer graphene}

\author{U Z\"ulicke$^{1,2}$ and R Winkler$^{3,4}$}

\address{$^1$ School of Chemical and Physical Sciences and
  MacDiarmid Institute for Advanced Materials and Nanotechnology,
  Victoria University of Wellington, PO Box 600, Wellington 6140,
  New Zealand}

\address{$^2$ Kavli Institute for Theoretical Physics, University of
  California, Santa Barbara, CA 93106, USA}

\address{$^3$ Department of Physics, Northern Illinois University,
  DeKalb, Illinois 60115, USA}

\address{$^4$ Materials Science Division, Argonne National
  Laboratory, Argonne, Illinois 60439, USA}

\ead{uli.zuelicke@vuw.ac.nz}

\begin{abstract}
  It was recently shown that a finite imbalance between electron
  densities in the $\vek{K}$ and $\vek{K}'$ valleys of bilayer
  graphene induces a magnetoelectric coupling. Here we explore
  ramifications of this electronically tunable magnetoelectric
  effect for the optical conductivity and dielectric permittivity of
  this material. Our results augment current understanding of
  longitudinal magnetoresistance and magnetocapacitance in
  unconventional materials.
\end{abstract}

\section{Introduction}
The coupling of electric and magnetic degrees of freedom in
materials continues to be studied intensely, because of both its
interesting physical origin and its potential for useful
applications~\cite{fie05, ram07, kho09, kim12, don15}. The recent
discovery~\cite{zue14, win15} of magnetoelectric couplings in
bilayer graphene (BLG) has established an intriguing link between
magnetoelectricity and atomically thin materials~\cite{nov05,
  xu13}. Here we investigate ramifications of this effect for
electric transport, pointing out intriguing similarities, but also
crucial differences, with transport characteristics~\cite{zyu12,
  gru12, son13, vaz13, bur15, gos15} arising from the chiral
anomaly~\cite{nie83} in Dirac~\cite{yan14} and Weyl~\cite{vaf14,
  has15} semimetals.

As a relevant context, we summarize the basic constitutive relations
that represent the chiral magnetic effect~\cite{nie83, zyu12, gru12,
  son13, vaz13, bur15, gos15} in Dirac and Weyl semimetals with two
valleys characterized by opposite chirality. The most important
feature is the generation of a finite imbalance $\varrho_\mathrm{v}$
of electron densities between the valleys when both an electric
field $\vekc{E}$ and a magnetic field $\vekc{B}$ are applied, which
also depends on the relaxation time $\tau_\mathrm{v}$ for
inter-valley scattering of charge carriers:
\begin{equation}\label{eq:WeylVCharge}
  \varrho_\mathrm{v} = -\frac{e^3}{2\pi^2\hbar^2}\,
  \tau_\mathrm{v}\, \vekc{B} \cdot \vekc{E} \quad .
\end{equation}
Combined with the fact that $\varrho_\mathrm{v}$ in the presence of
a magnetic field causes an electric current $\delta\vek{j}$ to flow,
\begin{equation}\label{eq:WeylCurr}
  \delta\vek{j} = -\frac{e}{2\pi^2\hbar^2\, D(E_\mathrm{F})}\,
  \vekc{B} \, \varrho_\mathrm{v} \quad ,
\end{equation}
a correction to the magnetoconductivity tensor emerges that is of the form
$\delta\ubar{\sigma}(\vekc{B}) = \delta\sigma(\mathcal{B}) \, \vekhc{B}
\vekhc{B}$, with
\begin{equation}
  \delta\sigma(\mathcal{B}) = \frac{e^4\,
    \tau_\mathrm{v}}{4\pi^4\hbar^4 \, D(E_\mathrm{F})} \,
  \mathcal{B}^2 \quad .
\end{equation}
Here $D(E_\mathrm{F})$ denotes the density of states at the Fermi
energy $E_\mathrm{F}$.  See, e.g., Refs.~\cite{zyu12, son13, bur15} for
relevant derivations and Ref.~\cite{xio15} for a recent experimental
observation.

As discussed in greater detail further below, the existence of a
magnetoelectric coupling in BLG whose magnitude depends on
$\varrho_\mathrm{v}$~\cite{zue14} gives rise to a set of
constitutive relations that are similar to those given above for
Dirac and Weyl semimetals.

\section{Magnetoelectric coupling in BLG for in-plane electric \& magnetic fields}
\label{sec:MEinBLG}
For the purpose of the present discussion, we only consider electric
fields $\vekc{E}_\|$ and magnetic fields $\vekc{B}_\|$
\emph{parallel to the plane\/} of the BLG sheet. (Effects of field
couplings in the general case are presented in Refs.~\cite{zue14,
  win15}.) The single-particle Hamiltonian describing the
charge-carrier dynamics is given by
\begin{equation}
  \label{eq:ham}
  H = H_0(\kk + e \vekc{A}) + e\, \xi_\|\, \vekc{E}_\| \cdot
  \vekc{B}_\|\, \tau_z \quad ,
\end{equation}
where $H_0(\kk)$ describes the low-energy band structure for the
field-free situation in the two valleys associated with the
high-symmetry $\vek{K}$ and $\vek{K}'$ points in the BLG Brillouin
zone, and $\vekc{A}$ is the electromagnetic vector potential. The
diagonal Pauli matrix $\tau_z$ operates in valley (isospin) space,
and $\xi_\|$ is a material-dependent parameter quantifying the
strength of magnetoelectric coupling.

Several unusual physical effects arise from the presence of the term
proportional to $\vekc{E}_\| \cdot\vekc{B}_\|\, \tau_z$. Due to its
asymmetric coupling to the $\vek{K}$ and $\vek{K}'$ valleys,
simultaneous application of electric and magnetic fields gives rise
to a valley-charge imbalance
\begin{equation}\label{eq:MErhoV}
  \varrho_\mathrm{v} \equiv \varrho^{\vek{K}} - \varrho^{\vek{K}'} =
  2 e^2\, \xi_\|\, D(E_\mathrm{F})\, \vekc{B}_\|\cdot \vekc{E}_\|
  \quad .
\end{equation}
This represents an intriguing analogy with the situation in the
Dirac or Weyl semimetals, see Eq.~(\ref{eq:WeylVCharge})
above. Furthermore, mirroring phenomena associated with axion
electrodynamics~\cite{wil87, fra08} or, more generally, with the
magnetoelectric effect~\cite{heh08, qi08, ess10}, the in-plane
magnetic field in conjunction with a finite $\varrho_\mathrm{v}$
induces an electric current~\cite{zue14}
\begin{equation}\label{eq:MEcurrent}
  \delta\vek{j} \equiv \vek{j}^{\vek{K}} + \vek{j}^{\vek{K}'} =
  \xi_\|\, \vekc{B}_\| \, \partial_t \varrho_\mathrm{v} \quad .
\end{equation}
While there is some similarity between (\ref{eq:MEcurrent}) and the
constitutive relation (\ref{eq:WeylCurr}) for the Weyl semimetals,
there is the crucial difference that the current arising from
magnetoelectric coupling in BLG is not proportional to
$\varrho_\mathrm{v}$ itself but to its time derivative.

\section{Magnetotransport from magnetoelectric coupling in BLG}
\label{sec:longDrude}
The main focus of the present work is to elucidate how
electric-transport characteristics of BLG are modified by the
magnetoelectric coupling. Hence, we assume $\vekc{B}_\|$ to be
static, but $\vekc{E}_\|$ to be time dependent, $\vekc{E}_\|(t)$.
Assuming furthermore spatial uniformity, the continuity equation for
$\varrho_\mathrm{v}$ becomes
\begin{equation}
  \partial_t\varrho_\mathrm{v} +
  \frac{\varrho_\mathrm{v}}{\tau_\mathrm{v}} = 2 e^2\, \xi_\| \,
  D(E_\mathrm{F})\, \vekc{B}_\|\cdot \partial_t \vekc{E}_\| \quad .
\end{equation}
Straightforward solution using Fourier transformation yields
\begin{equation}\label{eq:valleyCharge}
  \varrho_\mathrm{v}(\omega) = 2 e^2\, \xi_\|\, D(E_\mathrm{F})\,
  \vekc{B}_\|\cdot \vekc{E}_\|(\omega)\,\, \frac{1}{1 -
    i/(\omega\tau_\mathrm{v})} \quad .
\end{equation}
Inserting (\ref{eq:valleyCharge}) into the Fourier-transformed
Eq.~(\ref{eq:MEcurrent}), we obtain a contribution to the BLG-sheet
conductivity tensor that is due to the magnetoelectric coupling, $\delta
\ubar{\sigma}(\omega, \vekc{B}_\|) = \delta \sigma(\omega, \mathcal{B}_\|)
\,\, \vekhc{B}_\| \vekhc{B}_\|$, with
\begin{equation}
  \delta\sigma(\omega, \mathcal{B}_\|) = - 2 e^2\, \frac{D
    (E_\mathrm{F}) \, \xi_\|^2} {\tau_\mathrm{v}} \,
  \mathcal{B}_\|^2 \, \frac{(\omega\tau_\mathrm{v})^2}{\sqrt{1 +
      (\omega\tau_\mathrm{v})^2}} \, \ee^{-i
    \arctan(\omega\tau_\mathrm{v})}\quad .
\end{equation}
Thus the valley-density-dependent magnetoelectric coupling causes a
reduced conductivity in the direction parallel to an applied
in-plane magnetic field. The magnitude of the associated response is
set by the scale
\begin{equation}
  \sigma_{\mathrm{ME}}(\mathcal{B}_\|) = 2 e^2\, \frac{D
    (E_\mathrm{F}) \, \xi_\|^2} {\tau_\mathrm{v}} \,
  \mathcal{B}_\|^2 \quad ,
\end{equation}
which depends quadratically on $\mathcal{B}_\|$ but inversely on
$\tau_\mathrm{v}$. In the static limit $\omega\tau_\mathrm{v}\ll 1$,
the dissipative part of the conductivity dominates but is also
suppressed by a small factor $(\omega\tau_\mathrm{v})^2$. The
opposite (high-frequency) limit $\omega\tau_\mathrm{v}\gg 1$ is
characterized by the conductivity correction being dominantly
capacitive with magnitude proportional to the large factor
$\omega\tau_\mathrm{v}$, with the dissipative part saturating at
$\sigma_{\mathrm{ME}} (\mathcal{B}_\|)$.

\begin{figure}[b]
  \includegraphics[width=18pc]{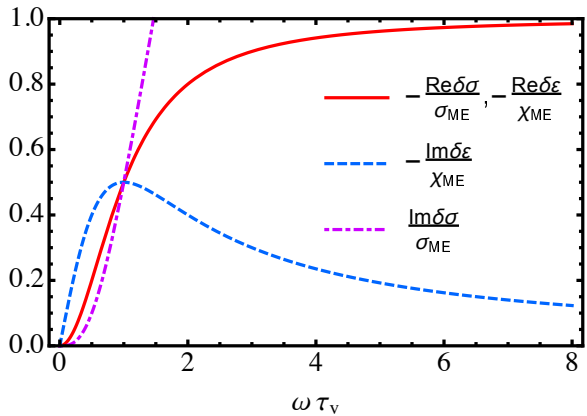}\hspace{\fill}%
  \begin{minipage}[b]{18pc}\caption{\label{fig:freqDeps}Frequency
      dependence of the real and imaginary parts of the contribution
      $\delta\sigma$ to the longitudinal sheet conductivity in BLG
      and the contribution $\delta \varepsilon$ to the dielectric
      constant arising due to the valley-density-dependent
      magnetoelectric effect. $\tau_\mathrm{v}$ is the inter-valley
      relaxation time.}
  \end{minipage}
\end{figure}

\section{Magnetodielectric effect in BLG}
\label{sec:magdielec}
Using the familiar identity relating the 3D-bulk conductivity and the
dielectric permittivity~\cite{jac99}, we find a magnetoelectric-coupling-induced
change to the dielectric tensor,
\begin{eqnarray}
  \frac{\delta \ubar{\varepsilon}(\omega,
    \vekc{B}_\|)}{\varepsilon_0} &\equiv& i
  \frac{\delta\ubar{\sigma}(\omega, \vekc{B}_\|)}{\varepsilon_0 \,
    \omega\, d} = \frac{\delta \varepsilon(\omega,
    \mathcal{B}_\|)}{\varepsilon_0} \,\, \vekhc{B}_\| \vekhc{B}_\|
  \quad , \\ \frac{\delta \varepsilon(\omega,
    \mathcal{B}_\|)}{\varepsilon_0} &=& - \frac{2 e^2}
        {\varepsilon_0} \, \frac{D(E_\mathrm{F})\, \xi_\|^2}{d}\,
        \mathcal{B}_\|^2 \, \frac{\omega\tau_\mathrm{v}}{\sqrt{1 +
            (\omega\tau_\mathrm{v})^2}} \, \ee^{i \arccot(\omega
          \tau_\mathrm{v})}\quad .
\end{eqnarray}
Here $d$ is the electronic width of BLG. Thus this material exhibits a negative
in-plane magnetodielectric effect~\cite{kim03,sin05} with a magnitude
that saturates at the susceptibility scale
\begin{equation}\label{eq:chiME}
  \chi_{\mathrm{ME}}(\mathcal{B}_\|) = \frac{2 e^2}{\varepsilon_0}
  \, \frac{D(E_\mathrm{F}) \, \xi_\|^2}{d}\, \mathcal{B}_\|^2
\end{equation}
in the high-frequency limit $\omega\tau_\mathrm{v}\gg 1$ but which
acquires a prefactor $\omega\tau_\mathrm{v}$, and therefore
vanishes, in the static limit $\omega\tau_\mathrm{v} \ll 1$. The
imaginary part of the dielectric function exhibits the lineshape of
the derivative of a Lorentzian, thus its measurement would allow
convenient extraction of the inter-valley relaxation time
$\tau_\mathrm{v}$. See Fig.~\ref{fig:freqDeps} for an illustration
of the frequency dependencies exhibited by the real and imaginary
parts of the contributions to the conductivity and the dielectric
constant due to the magnetoelectric effect.

\section{Discussion and conclusions}
\label{sec:concl}
Our study has revealed the existence of a longitudinal
magnetoconductivity and an associated magnetocapacitance in BLG
arising from its valley-density-dependent magnetoelectric effect.
More broadly, this behavior will be exhibited also by other
materials whose symmetries are similar to BLG. The ability to
generate a valley-density imbalance $\varrho_\mathrm{v}$ by
simultaneously applied non-orthogonal electric and magnetic fields
constitutes an intriguing analogy between BLG and the Dirac and Weyl
semimetals. Furthermore, in both material classes, a magnetic field
causes electric current to flow when the valley densities are
unequal, but this imbalance must also be time-varying in BLG whereas
a current is generated already by a static $\varrho_\mathrm{v}$ in
Dirac and Weyl semimetals. Thus, while the unusual magnetotransport
properties emerging in BLG and those associated with the chiral
anomaly in Dirac and Weyl semimetals turn out to share certain
features, they have a fundamentally different origin and also
exhibit different dependencies on physical parameters such as the
inter-valley relaxation time.

\ack UZ thanks M.~S.\ Fuhrer, I.\ Martin, and J.~E.\ Moore for useful
discussions and acknowledges support from NSF Grant No.\ PHY11-25915
while at the Kavli Institute for Theoretical Physics.  RW was
supported by the NSF under grant No.\ DMR-1310199.  Work at Argonne
was supported by DOE BES under Contract No.\ DE-AC02-06CH11357.

\section*{References}

\begin{thebibliography}{10}
\expandafter\ifx\csname url\endcsname\relax
  \def\url#1{{\tt #1}}\fi
\expandafter\ifx\csname urlprefix\endcsname\relax\def\urlprefix{URL }\fi
\providecommand{\eprint}[2][]{\url{#2}}

\bibitem{fie05}
Fiebig M 2005 {\em J. Phys. D\/} {\bf 38} R123--52

\bibitem{ram07}
Ramesh R and Spaldin N~A 2007 {\em Nat. Mater.\/} {\bf 6} 21--9

\bibitem{kho09}
Khomskii D 2009 {\em Physics\/} {\bf 2} 20

\bibitem{kim12}
Kimura T 2012 {\em Annu. Rev. Condens. Matter Phys.\/} {\bf 3} 93--110

\bibitem{don15}
Dong S, Liu J, Cheong S and Ren Z 2015 {\em Adv. Phys.\/} {\bf 64} 519--626

\bibitem{zue14}
Z\"ulicke U and Winkler R 2014 {\em Phys. Rev. B\/} {\bf 90} 125412

\bibitem{win15}
Winkler R and Z\"ulicke U 2015 {\em Phys. Rev. B\/} {\bf 91} 205312

\bibitem{nov05}
Novoselov K~S, Jiang D, Schedin F, Booth T~J, Khotkevich V~V, Morozov S~V and
  Geim A~K 2005 {\em Proc. Natl. Acad. Sci. USA\/} {\bf 102} 10451--53

\bibitem{xu13}
Xu M, Liang T, Shi M and Chen H 2013 {\em Chem. Rev.\/} {\bf 113} 3766--98

\bibitem{zyu12}
Zyuzin A~A and Burkov A~A 2012 {\em Phys. Rev. B\/} {\bf 86} 115133

\bibitem{gru12}
Grushin A~G 2012 {\em Phys. Rev. D\/} {\bf 86} 045001

\bibitem{son13}
Son D~T and Spivak B~Z 2013 {\em Phys. Rev. B\/} {\bf 88} 104412

\bibitem{vaz13}
Vazifeh M~M and Franz M 2013 {\em Phys. Rev. Lett.\/} {\bf 111} 027201

\bibitem{bur15}
Burkov A~A 2015 {\em Phys. Rev. B\/} {\bf 91} 245157

\bibitem{gos15}
Goswami P, Pixley J~H and Das~Sarma S 2015 {\em Phys. Rev. B\/} {\bf 92} 075205

\bibitem{nie83}
Nielsen H and Ninomiya M 1983 {\em Phys. Lett. B\/} {\bf 130} 389--96

\bibitem{yan14}
Yang B~J and Nagaosa N 2014 {\em Nat. Commun.\/} {\bf 5} 4898

\bibitem{vaf14}
Vafek O and Vishwanath A 2014 {\em Annu. Rev. Condens. Matter Phys.\/} {\bf 5}
  83--112

\bibitem{has15}
Hasan M~Z, Xu S~Y and Bian G 2015 {\em Phys. Scr.\/} {\bf T164} 014001

\bibitem{xio15}
Xiong J, Kushwaha S~K, Liang T, Krizan J~W, Hirschberger M, Wang W, Cava R~J
  and Ong N~P 2015 {\em Science\/} {\bf 350} 413--16

\bibitem{wil87}
Wilczek F 1987 {\em Phys. Rev. Lett.\/} {\bf 58} 1799--802

\bibitem{fra08}
Franz M 2008 {\em Physics\/} {\bf 1} 36

\bibitem{heh08}
Hehl F~W, Obukhov Y~N, Rivera J~P and Schmid H 2008 {\em Phys. Lett. A\/} {\bf
  372} 1141--46

\bibitem{qi08}
Qi X~L, Hughes T~L and Zhang S~C 2008 {\em Phys. Rev. B\/} {\bf 78} 195424

\bibitem{ess10}
Essin A~M, Turner A~M, Moore J~E and Vanderbilt D 2010 {\em Phys. Rev. B\/}
  {\bf 81} 205104

\bibitem{jac99}
Jackson J~D 1999 {\em Classical Electrodynamics\/} 3rd ed (Hoboken, NJ: Wiley)
  {S}ection 7.5 C

\bibitem{kim03}
Kimura T, Kawamoto S, Yamada I, Azuma M, Takano M and Tokura Y 2003 {\em Phys.
  Rev. B\/} {\bf 67} 180401

\bibitem{sin05}
Singh M~P, Prellier W, Simon C and Raveau B 2005 {\em Appl. Phys. Lett.\/} {\bf
  87} 022505

\end{thebibliography}
\providecommand{\newblock}{}

\end{document}